\documentclass[10pt,final]{article}
\pdfoutput=1

\usepackage{cite} 
\usepackage{fullpage}
\usepackage{url}
\usepackage{setspace}
\usepackage{comment}

\usepackage{subfig}
\usepackage{graphicx}

\title{DMTCP: Transparent Checkpointing for Cluster Computations and the Desktop}

\author{Jason Ansel\\
    Computer Science and\\
    \hbox{\ \ } Artificial Intelligence Laboratory\\
    MIT\\
    Cambridge, MA 02139 / USA \\
    jansel@csail.mit.edu
 \and
    \hbox{\ \ \ }Kapil Arya\hbox{\ \ \ \ \ \ \ }Gene Cooperman\thanks
              {The computational facilities for this work were
               partially supported by
               the National Science Foundation under Grant
               CNS-06-19616.} \\
    College of Computer\\
    \hbox{\ \ } and Information Sciences\\
    Northeastern University\\
    Boston, MA 02115 / USA \\
    {\{kapil,gene\}@ccs.neu.edu}
}
\date{}

\begin{document}

\maketitle

\begin{abstract}

DMTCP (Distributed MultiThreaded CheckPointing) is a transparent user-level
checkpointing package for distributed applications.  Checkpointing and restart
is demonstrated for a wide range of over 20~well known applications, including
MATLAB, Python, TightVNC, MPICH2, OpenMPI, and runCMS.  RunCMS runs as a
680~MB image in memory that includes 540~dynamic libraries, and is used for
the CMS experiment of the Large Hadron Collider at CERN.  DMTCP transparently
checkpoints general cluster computations consisting of many nodes, processes,
and threads; as well as typical desktop applications.  On 128~distributed
cores (32~nodes), checkpoint and restart times are typically 2~seconds,
with negligible run-time overhead.  Typical checkpoint times are reduced to
0.2~seconds when using forked checkpointing.  Experimental results show that
checkpoint time remains nearly constant as the number of nodes increases on
a medium-size cluster.

DMTCP automatically accounts for fork, exec, ssh, mutexes/semaphores, TCP/IP
sockets, UNIX domain sockets, pipes, ptys (pseudo-terminals), terminal modes,
ownership of controlling terminals, signal handlers, open file descriptors,
shared open file descriptors, I/O (including the readline library), shared
memory (via mmap), parent-child process relationships, pid virtualization,
and other operating system artifacts.  By emphasizing an unprivileged,
user-space approach, compatibility is maintained across Linux kernels from
2.6.9 through the current~2.6.28.  Since DMTCP is unprivileged and does not
require special kernel modules or kernel patches, DMTCP can be incorporated
and distributed as a checkpoint-restart module within some larger package.

\end{abstract}

\section{Introduction}

Checkpointing packages have been available for over 20~years.  They are now
often used in high performance computing and batch environments. Yet, they
have not widely penetrated to ordinary applications on the desktop. There is a
need for a simple transparent checkpointing package for commonly used desktop
applications (including binary-only commercial applications provided by a
third-party), that can at the same time handle distributed, multithreaded
production computations on clusters.  Unlike the most widely used systems
today, DMTCP is {\em user-level}, requiring no system privileges to operate.
This allows DMTCP to be bundled with the application, thereby opening up
entirely new applications for checkpointing.

As one striking example use case, many programs have a CPU-intensive first
phase, followed by a second phase for interactive analysis.  The approach
described here immediately enables a user to run the CPU-intensive portion of a
computation on a powerful computer or cluster, and then migrate the computation
to a single laptop for later interactive analysis at home or on a plane.

Another application of checkpointing that has been well received among users
of DMTCP is the ability to easily debug long-running jobs.  When bugs in the
middle of a long-running job are discovered, the programmer can repeatedly
restart from a checkpoint taken just before the bug occurred and examine
the bug in a debugger. This reduces the debug-recompile cycle for such cases.

Checkpointing is an inherently hard problem to solve in a robust way.
The most widely used checkpointing approaches today are based on custom
kernel modules.  This approach limits the applications of checkpointing
because it can only be deployed in controlled environments.  Additionally,
kernel modules are hard to maintain because they directly access internals
of the kernel that change more rapidly than standard APIs.  As evidence
of this, the web site checkpointing.org includes many earlier attempts at
checkpointing. Most of the attempts that were based on modification of the
kernel do not work on current kernel versions.

Checkpoint/restart has been successful to date in batch systems for production
use (but sometimes with restrictions on distributed or multithreaded
applications).  Here, it is reasonable to spend large amounts of manpower
to maintain kernel-specific checkpointing schemes.  This motivates why many
batch queues make checkpointing available, but in contrast, ``checkpointing on
the desktop'' is not as widely available today.  DMTCP tries to support both
traditional high performance applications and typical desktop applications.
With this in mind, DMTCP supports the critical feature of transparency:
no re-compilation and no re-linking of user binaries.  Because it supports
a wide range of recent Linux kernels (2.6.9 through the current 2.6.28), it
can be packaged as just one module in a larger application.  The application
binary needs no root privilege and need not be re-configured for new kernels.
This also painlessly adds a ``save/restore workspace'' capability to an
application, or even to a problem-solving environment with many threads
or processes.

Ultimately, the novelty of DMTCP rests on its particular combination
of features: user-level, multithreaded, distributed processes connected
with sockets, and fast checkpoint times with negligable overhead while not
checkpointing.  Those features are designed to support a broad range of use
cases for checkpointing, which go beyond the traditional use cases of today.

\subsection{Use Cases}
\label{sec:useCases}
In this section, we present some uses of checkpointing that go beyond the
traditional checkpointing of long-running batch processes.  Many of these
additional uses are motivated by desktop applications.

\begin{enumerate}
\item save/restore workspace: Interactive languages frequently include
  their own ``save/restore workspace'' commands.  DMTCP eliminates
  that need.
\item ``undump'' capability: programs that would otherwise have long
  startup times often create a custom ``dump/undump'' facility.  The
  software is then built, dumped after startup, and re-built to
  package a ``checkpoint'' along with an undump routine.  One of the
  applications for which we are working with the developers; cmsRun,
  has exactly this problem: initialization of 10 minutes to half an
  hour due to obtaining reasonably current data from a database, along
  with issues of linking approximately 400 dynamic libraries:
  unacceptable when many thousands of such runs are required.
\item a substitute for PRELINK: PRELINK is a Linux technology for
  prelinking an application, in order to save startup time when many
  large dynamic libraries are invoked.  PRELINK must be maintained
  in sync with the changing Linux architecture.
\item debugging of distributed applications: all processes are
  checkpointed just before a bug and then restarted (possibly on a single
  host) for debugging.
\item checkpointed image as the ``ultimate bug report''
\item applications with CPU-intensive front-end and interactive
  analysis of results at back-end: Run on high performance host or
  cluster, and restart all processes on a single laptop
\item traditional checkpointing of long-running distributed
  applications that may run under some dialect of MPI, or under a
  custom sockets package (e.g. iPython, used in SciPy/NumPy for
  parallel numerical applications.)
\item robustness: upon detecting distributed deadlock or race, automatically
  revert to an earlier checkpoint image and restart in slower, ``safe mode'',
  until beyond the danger point.
\end{enumerate}

\subsection{Outline}

Section~\ref{sec:relatedWork} covers related work.
Section~\ref{sec:usage} describes DMTCP as seen by an end-user.
Section~\ref{sec:architecture} describes the software
architecture.  
Section~\ref{sec:experiment} presents experimental results.
Finally, Section~\ref{sec:conclusion} presents the conclusions and future work.

\section{Related Work}
\label{sec:relatedWork}

There is a long history of checkpointing packages (kernel- and user-level,
coordinated and uncoordinated, single-threaded \hbox{vs.} multithreaded,
etc.).  Given the space limitations, we highlight only the most significant
of the many other approaches.

DejaVu~\cite{DejaVu} (whose development overlapped that of DMTCP) also provides
transparent user-level checkpointing of distributed process based on sockets.
However, DejaVu appears to be much slower than DMTCP.  For example, in the
Chombo benchmark, Ruscio \hbox{et al.} report executing ten checkpoints per
hour with 45\%~overhead.  In comparison, on a benchmark of similar scale DMTCP
typically checkpoints in 2~seconds, with essentially zero overhead between
checkpoints.  Nevertheless, DejaVu is also able to checkpoint InfiniBand
connections by using a customized version of MVAPICH.  DejaVu takes a more
invasive approach than DMTCP, by logging all communication and by using
page protection to detect modification of memory pages between checkpoints.
This accounts for additional overhead during normal program execution that
is not present in DMTCP.  Since DejaVu was not publicly available at the
time of this writing, a direct timing comparison on a common benchmark was
not possible.

The remaining work on distributed transparent checkpointing can be divided
into two categories:
\begin{enumerate}
\item {\em User-level MPI libraries for
    checkpointing}~\cite{Herault06,OpenMPICheckpoint07,CheckpointLAMMPI05,CheckpointFTC-Charm,CoCheck,MPICheckpoint,MPICH-GF,MPIfaultTolerant,MPICheckpointing}:
  works for distributed processes, but only if they communicate
  exclusively through MPI (Message Passing Interface).  Typically
  restricted to a particular dialect of MPI.
\item {\em Kernel-level (system-level)
    checkpointing}~\cite{BLCR,CheckpointLAMMPI05b,Cruz05,LaadanEtAl05,LaadanEtAl07,CheckpointSMP,Chpox}:
  modification of kernel; requirements on matching package
  version to kernel version. 
\end{enumerate}

A crossover between these two categories is the kernel level checkpointer
BLCR~\cite{BLCR,CheckpointLAMMPI05b}.  BLCR is particularly notable because of
its widespread usage.  BLCR itself can only checkpoint processes on a single
machine.  However some MPI libraries (including some versions of OpenMPI,
LAM/MPI, MVAPICH2, and MPICH-V) are able to integrate with BLCR to provide
distributed checkpointing.

Three notable distributed kernel-level solutions based
on the Linux kernel module Zap are provided by Laadan
and Nieh~\cite{LaadanEtAl05,LaadanEtAl07} and Janakiraman
et~{al.}~\cite{Cruz05}, and Chpox by Sudakov et~\hbox{al.}~\cite{Chpox}.
This approach leads to checkpoints being more tightly coupled to
kernel versions.  It also makes future ports to other operating
systems more difficult.

Much MPI-specific work has been based on {\em coordinated
checkpointing} and the use of hooks into communication by the MPI
library~\cite{Herault06,OpenMPICheckpoint07}.  In contrast,
our goal is to support more general distributed scientific software.

In addition to distributed checkpointing, many
packages exist which perform single-process
checkpointing~\cite{CheckpointLibckpt,DieterLumpp01,PinheiroEPCKPT,IncrementalCheckpoint95,ForkedCheckpointing90,ForkedCheckpointing94,Winckp99,abdelshafiefficient,CheckpointIncremental,CheckpointCondor}.

For completeness, we also note the complementary technology of virtual
machines.  As one example, VMware offers both snapshot and record/replay
technology for its virtual machines.  The process-level checkpointing of DMTCP
is inherently a lighter weight solution.  Further, process-level checkpointing
makes it easier to support distributed applications.  VMware players require
system privilege for installation, although snapshot and record/replay can
thereafter be used at user level.

Further discussion of checkpointing is found in the following
surveys~\cite{Survey0,Survey1,Survey2}.

\section{Usage and Features}
\label{sec:usage}

The user will typically use three DMTCP commands:
{\small
\begin{quotation}
\noindent
{\tt dmtcp\_checkpoint [options] <program>} \\
{\tt dmtcp\_command <command>} \\
{\tt dmtcp\_restart\_script.sh} 
\end{quotation}
}
The restart script is generated at checkpoint time.  Each invocation of
{\tt dmtcp\_checkpoint} by the end user causes the corresponding process
to be registered as one of the set of processes that will be checkpointed.
All local and remote child processes are checkpointed recursively.  As an
example, to run an MPICH-2 computation under DMTCP the user would first run:

{\small
\begin{quotation}
\noindent
{\tt dmtcp\_checkpoint mpdboot -n 32} \\
{\tt dmtcp\_checkpoint mpirun <mpi-program>}
\end{quotation}
}

Note that the MPI resource management processes are also checkpointed.
The first call to {\tt dmtcp\_checkpoint} will automatically spawn the
checkpoint coordinator.  {\tt mpdboot} will call {\tt ssh} to spawn remote
processes, these calls are transparently intercepted and modified so the
remote processes are also run under DMTCP. To request a checkpoint, the user
would then run:
\begin{quotation}
\noindent
{\tt\small dmtcp\_command --checkpoint} 
\end{quotation}

Checkpoints may also be generated at regular intervals by using the {\tt
--interval} option or requested by the application via the DMTCP programming
interface.  The checkpoint images for each process are written to unique
filenames in a user specified directory.  Additionally, a shell script, {\tt
dmtcp\_restart\_script.sh}, is created containing all the commands needed
to restart the distributed computation.  This script consists of many calls
to {\tt dmtcp\_restart}, one for each node.

A more detailed list of options and commands for controlling the behavior
of DMTCP are described in the manpages shipped with DMTCP.

\subsection{Programming Interface}

DMTCP is able to checkpoint unmodified Linux executables.  We envision
the typical use case as having the checkpointed application completely
unaware of DMTCP.  (This is the configuration used in experimental
results.)  However, for those wishing to have more control over the
checkpointing process, we provide a library for interacting with DMTCP
called {\tt dmtcpaware.a}.  This library allows the application to:
test if it is running under DMTCP; request checkpoints; delay checkpoints
during a critical section of code; query DMTCP status; and insert hook
functions before/after checkpointing or restart.

\section{Software Architecture}
\label{sec:architecture}

DMTCP consists of 17,000 lines of C and C++~code.  DMTCP is freely
available as open source software and can be downloaded from:
\begin{itemize}
\item[]\url{http://dmtcp.sourceforge.net/}
\end{itemize}
DMTCP is built upon our previous work, MTCP (MultiThreaded
CheckPointing)~\cite{RiekerAnselCooperman06}.  MTCP is assigned responsibility
for checkpointing of individual processes, while DMTCP checkpoints and
restores socket/file descriptors and other artifacts of distributed software.
This novel two-layer design greatly aids in maintenance and portability.

\subsection{Design of DMTCP}

DMTCP refers both to the entire package, and to the distributed
layer of the package.
The two layers of DMTCP, known as DMTCP and MTCP, consist of:

\begin{enumerate}

\item DMTCP allows checkpointing of a network
of processes spread over many nodes.  After DMTCP copies all inter-process
information to user space, it delegates single-process checkpointing to
a separate checkpoint package.

\item We base single-process checkpointing on our previous work, MTCP
 (MultiThreaded CheckPointing)~\cite{RiekerAnselCooperman06}.

\end{enumerate}

These two layers are separate, with a small API between
them.  This two-layer user-level approach has a potential advantage in
non-Linux operating systems, where DMTCP can be ported to run over other
single-process checkpointing packages that may already exist.

Checkpointing is added to arbitrary applications by injecting a shared
library at execution time.  This library:
\begin{itemize}
\item Launches a checkpoint management thread in every user process
which coordinates checkpointing.
\item Adds wrappers around a small number
of {\tt libc} functions in order to record information about open sockets
at their creation time. 
\end{itemize}
System calls and the {\tt proc} filesystem are also used to probe
kernel state. 

We use a {\em coordinated checkpointing} method, where all processes and
threads cluster-wide are simultaneously suspended during checkpointing.
Network data ``on the wire'' and in kernel buffers is flushed into the
recipient process's memory and saved in its checkpoint image. After a
checkpoint or restart, this network data is sent back to the original
sender and retransmitted prior to resuming user threads. A more detailed
account of our methodology can be found in Section~\ref{sec:architecture}

The only global communication primitive used at checkpoint time is a
barrier.  At restart time, we additionally require a discovery service
to discover the new addresses for processes migrated to new hosts.

\subsection{Initialization of an application process under DMTCP}
\label{sec:initialization}

At startup of a new process {\tt dmtcp\_checkpoint} injects
$dmtcphijack.so$, the DMTCP library responsible for checkpointing,
into the user program.  Library injection is currently done using
{\tt LD\_PRELOAD}.  Library injection can also be done after program
startup~\cite{Zandy99EtAl} and on other architectures~\cite{DLLInjection}.

Once injected into the user process, DMTCP loads $mtcp.so$, our single process
checkpointer, and calls the MTCP setup routines to enable integration with
DMTCP.  MTCP creates the checkpoint manager thread in this setup routine.
DMTCP also opens a TCP/IP connection to the checkpoint coordinator at this
time.  This results in a copy of our libraries and manager residing within
each checkpointed process.

DMTCP adds wrappers around a small number of {\tt libc} functions.  This is
done by overriding libc symbols with our library.  For efficiency reasons,
we avoid wrapping any frequently invoked system calls such as {\tt read}
and {\tt write}.  The wrappers are necessary since DMTCP must be aware of
all forked child processes, of all attempts to create remote processes (for
example via an exec to an ssh process), and of the parameters by which all
sockets are created.  In the case of sockets, DMTCP needs to know whether the
sockets are TCP/IP sockets (and whether they are listener or non-listener
sockets), UNIX domain sockets, or pseudo-terminals.  DMTCP places wrappers
around the following functions: socket, connect, bind, listen, accept,
setsockopt, fexecve, execve, execv, execvp, fork, close, dup2, socketpair,
openlog, syslog, closelog, ptsname and ptsname\_r.  The rest of this section
describes the purposes for these wrapper.

\subsection{Checkpointing under DMTCP}
\label{sec:checkpoint}

Checkpointing proceeds through seven stages and six global barriers.
Global barriers could be implemented efficiently through peer-to-peer
communication or broadcast trees, but are currently centralized for
simplicity of implementation.

\begin{figure*}[htb]
\begin{center}
\resizebox{!}{3in}
{\includegraphics{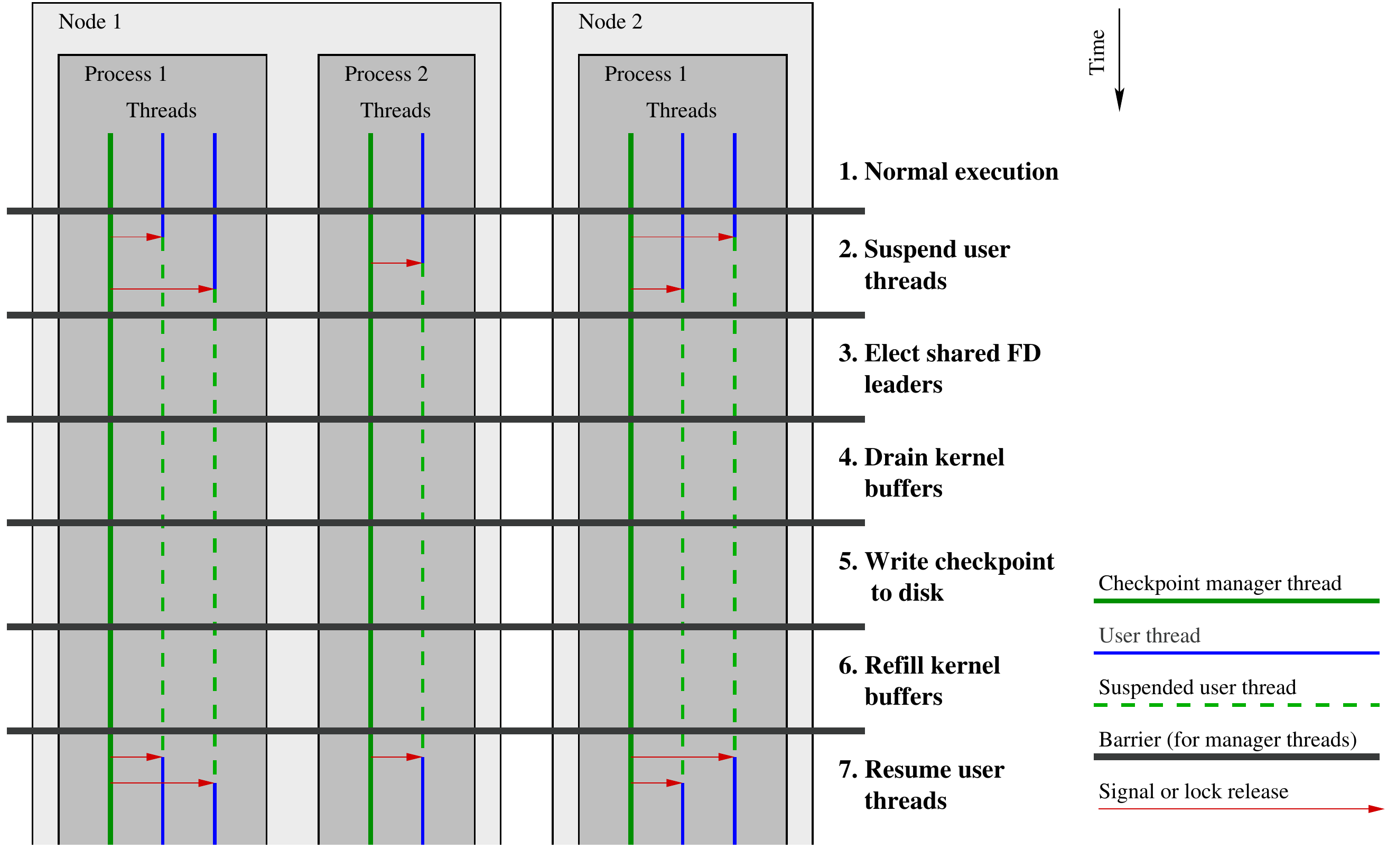}}
\end{center}
\caption{\label{fig:ckptdiag}Steps for checkpointing a simple system
with 2 nodes, 3 processes, and 5 threads.}
\end{figure*}

The following is the DMTCP distributed algorithm for checkpointing an entire
cluster.  It is executed asynchronously in each user process.  The only
communication primitive used is a cluster-wide barrier.  The following steps
are depicted graphically in Figure~\ref{fig:ckptdiag}.

\begin{enumerate}
  
  \item\label{step:ckptNormal}{\em Normal execution:}
    The checkpoint manager thread in each process waits until a new
    checkpoint is requested by the coordinator.  This is done by waiting
    at a special barrier that is not released until checkpoint time.
  
  \item\label{step:ckptSuspend}{\em Suspend user threads:}
     MTCP suspends all user threads, then DMTCP saves the owner of each file descriptor.
     DMTCP then waits until all application processes reach
     Barrier~2:  ``suspended'', then releases the barrier.
  
  \item\label{step:ckptElect}{\em Elect shard FD leaders:}
    DMTCP executes an election of a leader for each potentially
          shared file descriptor.  We trick the operating system into
          electing a leader for us by misusing the $F\_SETOWN$ flag of
          {\tt fcntl}.  All processes set the owner, and the last one
          wins the election.  In Step~\ref{step:ckptDrain}, a process can
          test if it is the election leader for a socket~{\tt fd} by testing
          if {\tt fcntl(fd, F\_GETOWN)==getpid()}.  The original value for
          $F\_SETOWN$ is restored after kernel buffers are refilled.
    DMTCP then waits until all application processes reach Barrier~3:
           ``election completed'', then releases the barrier.
  
  \item\label{step:ckptDrain}{\em Drain kernel buffers and perform handshakes with peers:}
     For each socket, the corresponding election leader flushes that socket
     by sending a special token.  It then drains that socket by receiving
     until there is no more available data and the special token is seen.
     DMTCP then performs handshakes with all socket peers to discover the
     {\em globally unique ID} of the remote side of all sockets.  The {\em
     connection information table} is then written to disk.  DMTCP then
     waits until all application processes reach Barrier~4:
      ``drained'', then releases the barrier.
  
  \item\label{step:ckptWrite}{\em Write checkpoint to disk:}
     The contents of all socket buffers is now in user space.
     MTCP writes all of user space memory to the checkpoint file.
     DMTCP then wait until all application processes reach Barrier~5:
      ``checkpointed'', then release the barrier.
  
  \item\label{step:ckptRefill}{\em Refill kernel buffers:}
     DMTCP then sends the drained socket buffer data back to the sender.
     The sender refills the kernel socket buffers by resending the data.
     DMTCP then waits until all application processes reach Barrier~6:
      ``refilled'', then releases the barrier.

  \item\label{step:ckptResume}{\em Resume user threads:}
    MTCP then resumes the application threads and
    DMTCP returns to Step~1.

\end{enumerate}

\subsection{Restart under DMTCP}
\label{sec:restart}

\begin{figure*}[htb]
  \begin{center}
  \resizebox{!}{3in}
  {\includegraphics{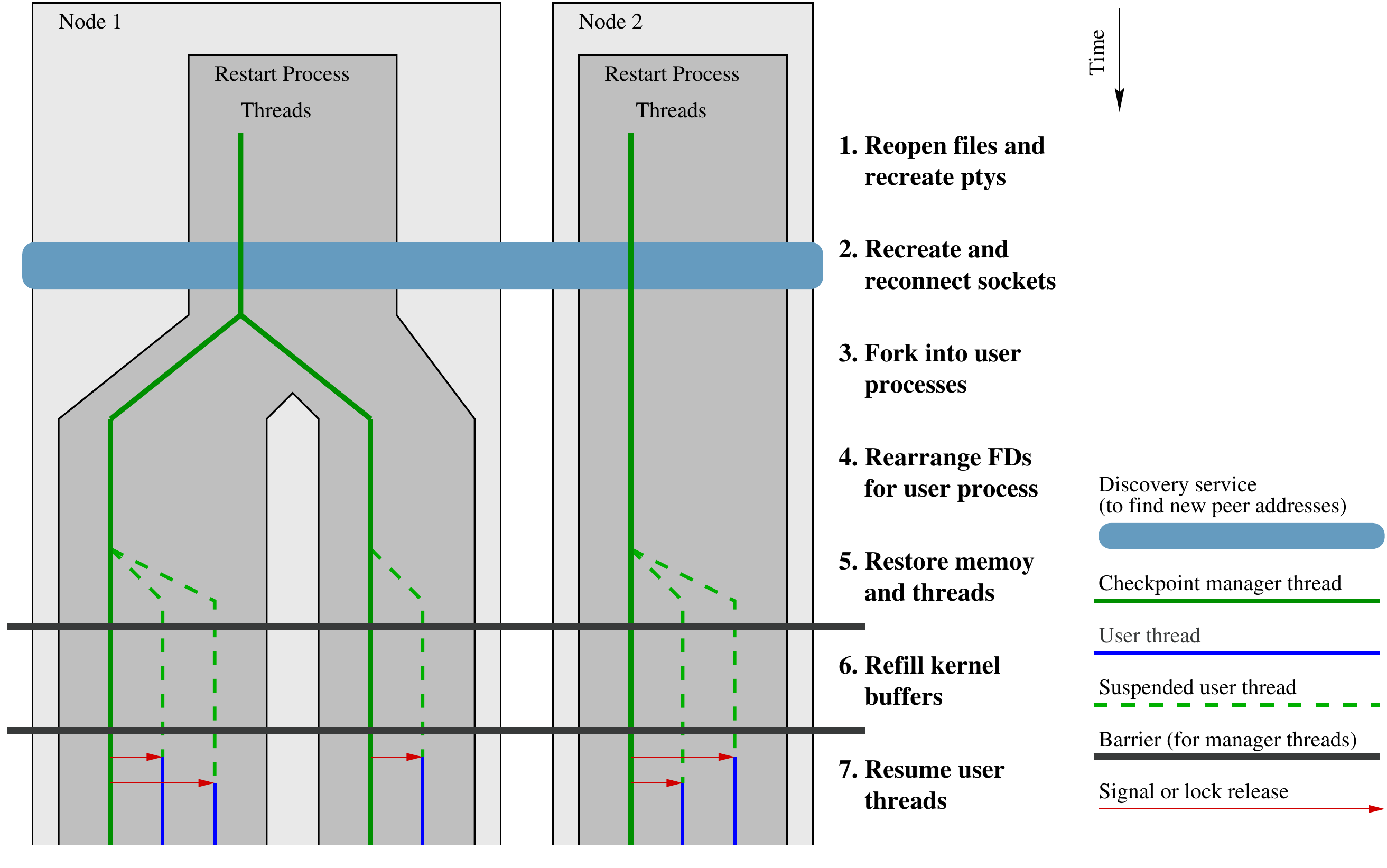}}
  \end{center}
  \caption{\label{fig:rstrdiag}
  Steps for restarting the system checkpointed in Figure~\ref{fig:ckptdiag}.
  The unified restart process and subsequent fork are required to recreate
  sockets and pipes shared between processes.
  }
\end{figure*}

The restart process undergoes some complexity in order to restore shared
sockets.  Under UNIX semantics multiple processes may share a single
socket connection.  When a process forks all open file descriptors become
shared between the child and parent.  To handle this, we refer to sockets
by a {\em globally unique ID} (hostid, pid, timestamp, per-process
connection number) and thus can detect duplicates at restart time.
These globally unique socket IDs (and other meta information), were
recorded at checkpoint time in the {\em connection information table}
for each process.  To recreated shared sockets, a {\em single} DMTCP
restart process is created on each host.  This single restart process
will first restore all sockets, and then fork to create each individual
user process on that host.

The following algorithm restarts the checkpointed cluster computation. 
It is executed asynchronously on each host in the cluster.  The steps of this
algorithm are depicted graphically in Figure~\ref{fig:rstrdiag}.
\begin{enumerate}
  
  \item\label{step:rstrFiles}{\em Reopen files and recreate ptys:}
    File descriptors, excluding sockets connected to a remote
    process, are regenerated first.  These include files, listen sockets,
    uninitialized sockets, and pseudo-terminals.
  
  \item\label{step:rstrSockets}{\em Recreate and reconnect sockets:}
    For each socket, the restart program uses the cluster-wide
    discovery service to find the new address of the corresponding
    restart process.  Once the new addresses are found the connections
    are re-established. The discovery services is needed since processes
    may be relocated between checkpoint and restore.
  
  \item\label{step:rstrFork}{\em Fork into user processes:}
    The DMTCP restart program now forks into N processes, where N is
    the number of user processes it intends to restore.

  \item\label{step:rstrRearrange}{\em Rearrange FDs for user process:}
    Each of these processes uses {\tt dup2} and {\tt close} to
      re-arrange the file descriptors to reflect the arrangement prior
      to checkpointing.  Unneeded file descriptors belonging to other
      processes are closed.  Shared file descriptors will now exist in
      multiple processes.
  
  \item\label{step:rstrMtcp}{\em Restore memory and threads:}
    The MTCP restart routine is now called to restore the local process
      memory and threads.  Upon completion the user process will resume
      at Barrier~5 of the checkpoint algorithm in Section~\ref{sec:checkpoint}.
  
  \item\label{step:rstrRefill}{\em Refill kernel buffers:}
    The program resumes as if it had just finished writing the
    checkpoint to disk, in Step~\ref{step:ckptRefill} of checkpointing.
  
  \item\label{step:rstrResume}{\em Resume user threads:}
    The program continues executing Step~\ref{step:ckptResume}
    of checkpointing.
\end{enumerate}

Step~\ref{step:rstrSockets} above bears further explanation. Recall
that prior to checkpointing, whenever a new connection was accepted,
wrappers around the system calls {\tt connect} and {\tt accept} had
transferred information about the {\tt connect}or to the {\tt accept}or.
This information includes a globally unique socket ID that remains
constant even if processes are relocated.  

At restart time, the {\tt accept}or for each socket advertises the
address and port of its restart lister socket to the discovery service.
When the {\tt connect}or receives this advertisement, it opens a new
connection to the {\tt accept}or who sent the advertisement.  The two
sides then perform a handshake and agree on the socket being restored.
Finally, {\tt dup2} is used on each side to move the socket descriptor
to the correct location.  This process continues asynchronously until
all sockets are restored.  Our methodology supports both sides of a
socket migrating.  It also supports loopback sockets.

\subsection{Implementation Strategies}
\label{sec:implementation}

\begin{figure*}[htb]
  \centering
    \subfloat[Checkpoint/Restart timings.]{\label{fig:realapps-time}
      \resizebox{0.9\linewidth}{!}{
        \includegraphics[angle=90]{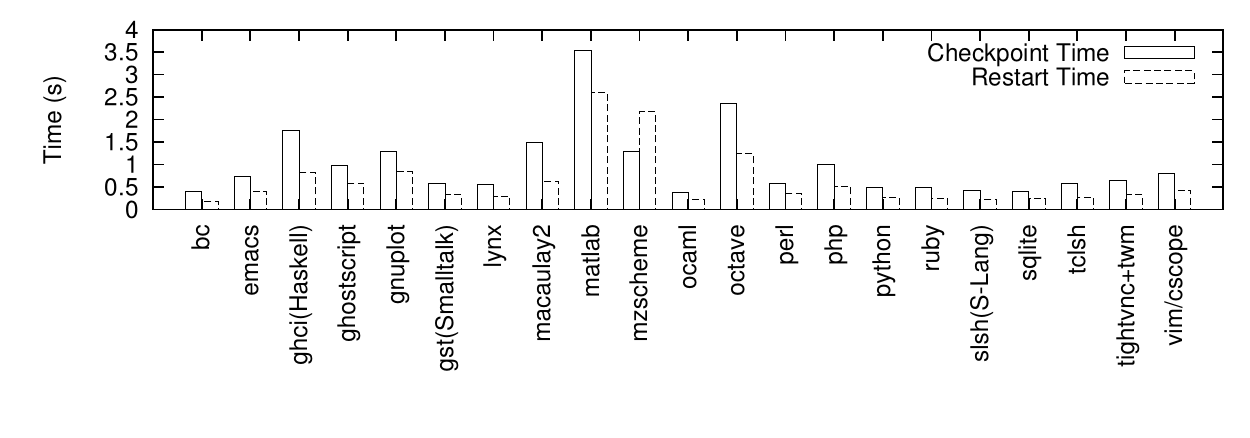}}}

    \subfloat[Checkpoint sizes.]{\label{fig:realapps-size}
      \resizebox{0.9\linewidth}{!}{
        \includegraphics[angle=90]{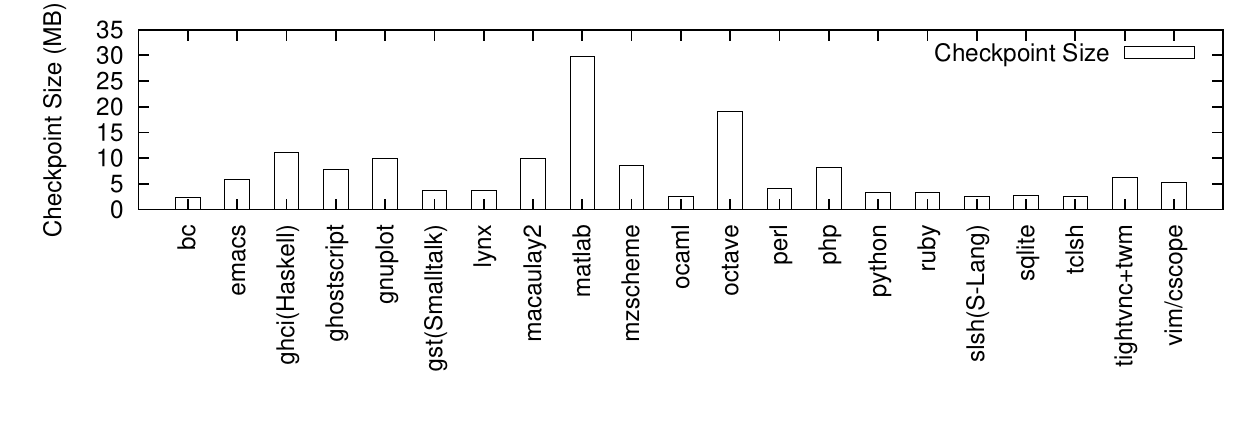}}}

  \caption{{\bf Common shell-like languages} and other applications. All
             are run on a single node with compression enabled.}

  \label{fig:realapps}
\end{figure*}

In the implementation, some less obvious issues arise in the support
for pipes, shared memory (via mmap), and virtual pids.

Pipes present an issue because they are unidirectional.  As seen
in Sections~\ref{sec:checkpoint} and~\ref{sec:restart}, the strategy
for checkpointing network data in a socket connection is for the
receiver to drain the socket into user space, then write a checkpoint
image, and finally re-send the network data through the same socket
back to the sender.  In order to support pipes, a wrapper around
the pipe system call promotes pipes into sockets.

In the case of shared memory, if the backing file of a shared memory
segment is missing and we have directory write permision, then we create
a new backing file.  Next, assuming the backing file is present and we
have write access, we overwrite the shared memory segment with
data from the checkpoint image.  If two processes share this memory,
they will both write to the same shared segment, but with the same data,
since the segment was also shared at the time of checkpoint.

If we do not have write access (for example, read-only access to certain
system-wide data), then we map the memory segment by the current data
of the file, and not the checkpoint image data.

In order to support virtual pids (process ids), one must worry about
pid conflicts.  The original pid When a process is first created through
a call to fork, its pid also becomes its virtual pid, and that virtual
pid is maintained throughout succeeding generations of restarts.  Hence,
a new process may have pid~A.  After checkpoint and restart, a second
process may be created with the same pid~A.  Our wrapper around fork
detects this situation, terminates the child with the conflicting virtual
pid,, and forks once again.

\section{Experimental Results}
\label{sec:experiment}

DMTCP is currently implemented for GNU/Linux.  The software has been
verified to work on recent versions of Ubuntu, Debian, OpenSuse, and
Red Hat Linux with Linux kernels ranging from version~2.6.9 through
version~2.6.28. DMTCP runs on x86, x86\_64 and mixed (32-bit processes
in 64-bit Linux) architectures.

Experiments were run on two broad classes of programs: shell-like languages
intended for a single computer (\hbox{e.g.} MATLAB, Perl, Python, Octave,
etc.); and distributed programs across the nodes of a cluster (\hbox{e.g.}
ParGeant4, iPython, MPICH2, OpenMPI, etc).  Reported checkpoint images are
after gzip compression (unless otherwise indicated), since DMTCP dynamically
invokes gzip before saving, by default.

In Section~\ref{sec:realworld}, our goal was to demonstrate on 20~common
real-world applications.  An emphasis on shell-like languages were chosen for
their widespread usage, and for their tendency to invoke multiple processes and
multiple threads in their implementation.  The languages were chosen from the
applications listed under ``Interactive mode languages'' (shell-like languages)
in the article ``List of programming languages by category'' on Wikipedia.

Section~\ref{sec:distributed} is concerned with testing for scalability.
The parallel tools and benchmarks were chosen for their popularity in
the computational science community.  They were augmented with some
computational packages that had already been configured and installed as tools
used by our own working group.

\subsection{Common Shell-Like Languages}
\label{sec:realworld}

These tests were conducted on a dual-socket, quad-core (8~total cores)
Xeon~E5320.  This system was running 64-bit Debian ``sid'' GNU/Linux
with kernel version~2.6.28.

To show breadth, we present checkpoint times, restart times, and checkpoint
sizes on a wide variety of commonly used applications.  These results are
shown in Figure~\ref{fig:realapps}.
These applications are:
BC (1.06.94) -- an arbitrary precision calculator language;  
Emacs (2.22) -- a well known text editor;
GHCi (6.8.2) -- the Glasgow Haskell Compiler;
Ghostscript (8.62) -- PostScript and PDF language interpreter;
GNUPlot (4.2) -- an interactive plotting program;
GST (3.0.3) -- the GNU Smalltalk virtual machine;
Lynx (2.8.7) -- a command line web browser;
Macaulay2 (2-1.1) -- a system supporting research in algebraic geometry and commutative algebra;
MATLAB (7.4.0) -- a high-level language and interactive environment for technical computing;
MZScheme (4.0.1) -- the PLT Scheme implementation;
OCaml (3.10.2) -- the Objective Caml interactive shell;
Octave (3.0.1) -- a high-level interactive language for numerical computations;
PERL (5.10.0) -- Practical Extraction and Report Language interpreter;
PHP (5.2.6) -- an HTML-embedded scripting language;
Python (2.5.2) -- an interpreted, interactive, object-oriented programming language;
Ruby (1.8.7) -- an interpreted object-oriented scripting language;
SLSH (0.8.2) -- an interpreter for S-Lang scripts;
SQLite (2.8.17) -- a command line interface for the SQLite database;
tclsh (8.4.19) -- a simple shell containing the Tcl interpreter;
TightVNC+TWM (1.3.9, 1.0.4) -- a headless X11 server running Tab Window Manager underneath it;
and
vim/cscope (15.6) -- interactively examine a C program.

Of particular interest is the checkpointing of TightVNC, a headless
X11 server.  We checkpoint the vncserver, the window manager, and all
graphical applications.  Between checkpoints, clients can connect with
(uncheckpointed) vncviewers to interact with the graphical applications.
Using this technique, we can checkpoint graphical applications without the
need to checkpoint interactions with graphics hardware.

Additionally (not included in the graphs, because of differences in scale) we
have demonstrated checkpointing of RunCMS.  RunCMS checkpoints in 25.2 seconds
and restarts in 18.4 seconds.  RunCMS is of especially timely interest, with
the recent startup of the large hadron collider at CERN.  We are collaborating
with the CMS experiment at CERN to checkpoint and restart their CMS software
of up to two million lines of code and up to 700~dynamic libraries.  We test
on a configuration which grows to 680~MB of data after running for 12~minutes.
At that time, it had loaded 540~dynamic libraries, as measured by the maps
file of the proc filesystem.  The checkpointed image file on disk was 225~MB,
after gzip compression.  (DMTCP invokes gzip compression by default.)

\subsection{Distributed Applications}
\label{sec:distributed}

\begin{figure*}[t]
  \centering
  \subfloat[Checkpoint timings.]{\label{fig:apps32cp}
    \resizebox{.48\linewidth}{!}{
      \includegraphics[angle=90]{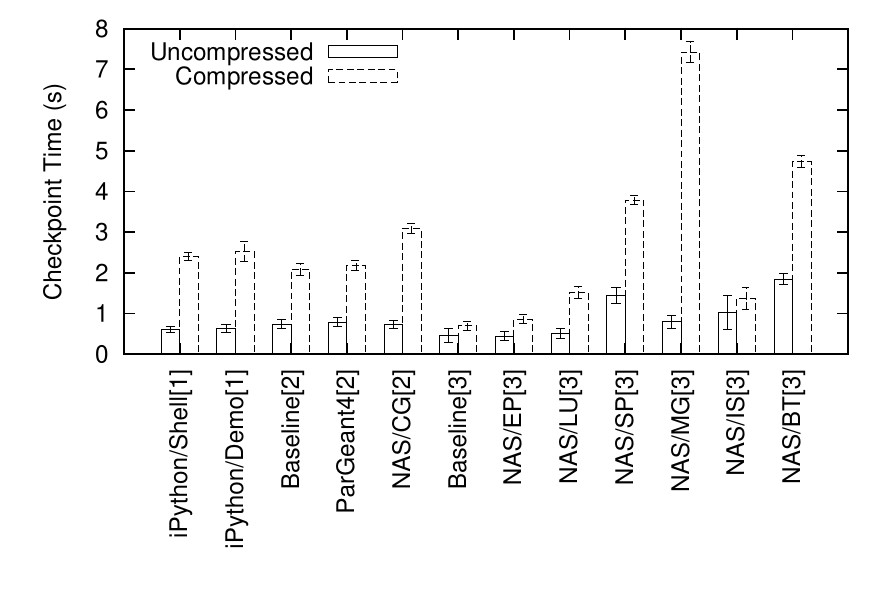}}}
  \subfloat[Restart timings.]{\label{fig:apps32rs}
    \resizebox{.48\linewidth}{!}{
      \includegraphics[angle=90]{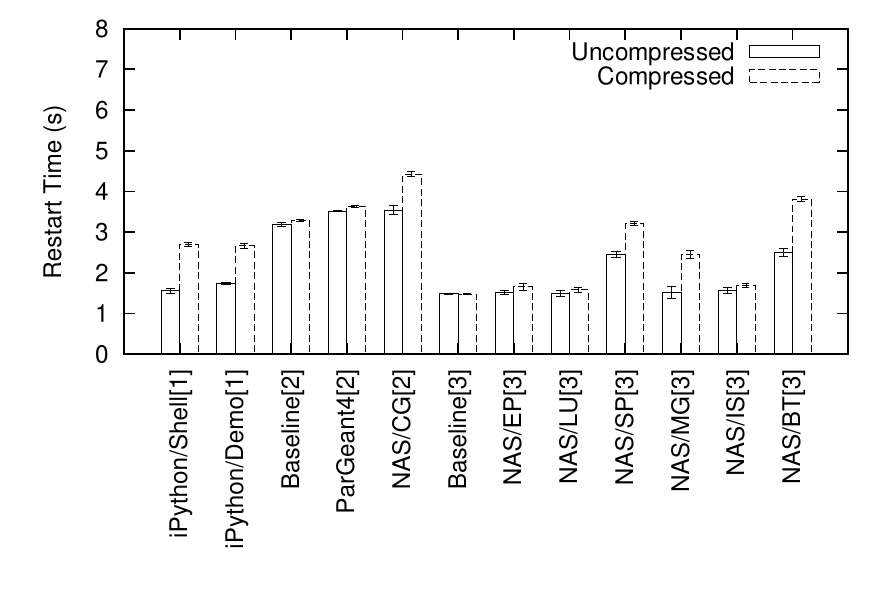}}}

  \subfloat[Aggregate (cluster-wide) checkpoint size.]{\label{fig:apps32sz}
    \resizebox{.5\linewidth}{!}{
      \includegraphics[angle=90]{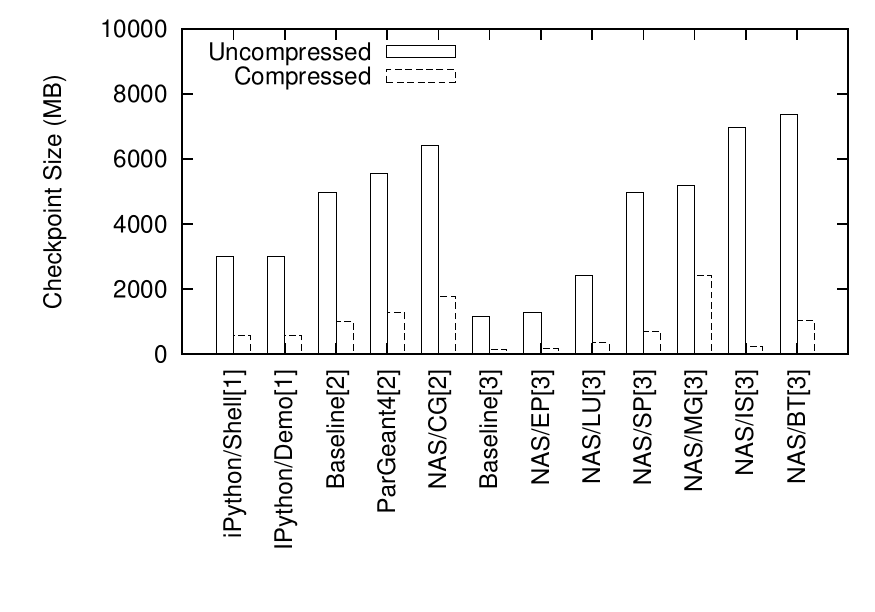}}}

  \caption{{\bf Distributed Applications.} Timings on 32 nodes.  Applications marked [1] use sockets
  directly.  Applications marked [2] are run using MPICH2.  Applications
  marked [3] are run using OpenMPI. Timing tests repeated 10 times, and the
  mean value is shown. Error bars in timings indicate plus or minus one
  standard deviation.}

  \label{fig:apps32}
\end{figure*}

\begin{figure}[ht]
  \noindent
  \centering
    \subfloat[Checkpoints stored to local disk of each node.]
     {\label{fig:nodespg4local}
      \resizebox{.48\linewidth}{!}{
       \includegraphics[width=.5\linewidth]{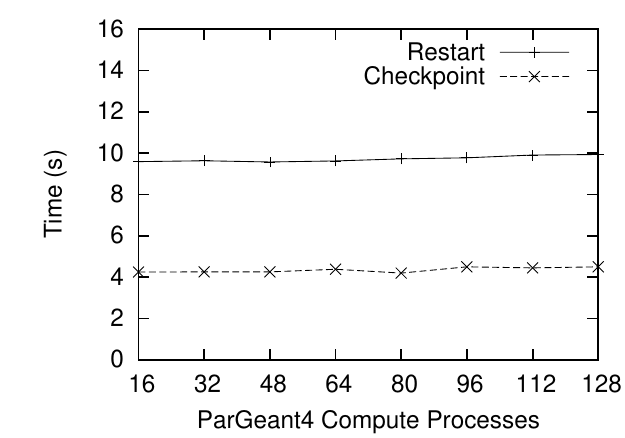}}}
    \subfloat[Checkpoints stored to centralized RAID storage via SAN and NFS.]
      {\label{fig:nodespg4san}
       \resizebox{.48\linewidth}{!}{
        \includegraphics[width=.5\linewidth]{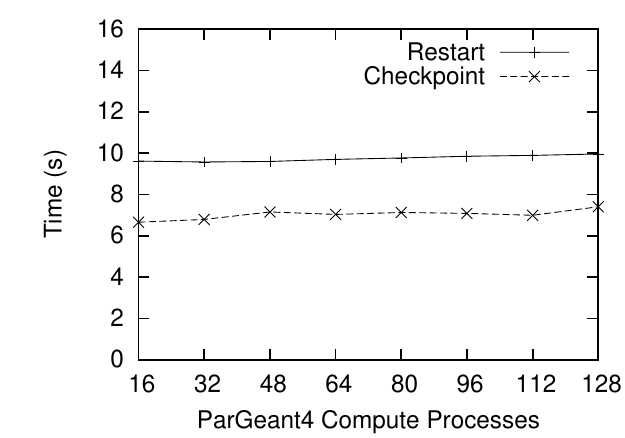}}}
    \caption{\label{fig:nodespg4}
      Timings as the number of processes and nodes changes.
      Application is ParGeant4 running under MPICH2.  Compression is
      enabled.  Compute processes per core and per node are held constant
      at 1 and~4, while the number of nodes is varied.  (Note: An additional 21
      to 161 MPICH2 resource management processes are also checkpointed.)}
\end{figure}

Distributed tests (Section~\ref{sec:distributed}) were conducted
on a 32~node cluster with 4~cores per node (128 total cores).  Each node was
configured with dual-socket,dual-core Xeon~5130 processors and 8~GB or 16~GB
of RAM.  Each node was running 64-bit Red Hat Enterprise GNU/Linux release~4
with kernel version~2.6.9-34.  The cluster was connected with Gigabit Ethernet.

In Figure~\ref{fig:nodespg4san}, checkpoints were written to a centralized
EMC CX300 SAN storage device over a 4 Gbps Fibre Channel Switch.
(SAN stands for storage area network.) On our cluster, only 8 of the 32
nodes were connected to the SAN. The remaining 24 nodes wrote indirectly
to the storage device via NFS.  In all other tests, checkpoints were
written to local disk of each node.

We report checkpoint times, restart times, and checkpoint file sizes
for a suite of distributed applications.  These results are contained in
Figures~\ref{fig:apps32cp}, \ref{fig:apps32rs} and~\ref{fig:apps32sz},
respectively.  In each case, we report the timings and file sizes both with
and without compression. The following applications are shown:

\begin{itemize}
\item
{\bf Based on sockets directly:}
\begin{itemize}
\item {iPython:}~\cite{ipython} an enhanced Python shell with support for parallel and distributed computations.  Used in scientific computations such as SciPy/NumPy.  {\tt iPython/Shell:} is the interactive iPython interpreter, idle at time of checkpoint.
{\tt iPython/Demo:} is the ``parallel computing'' demo included with the iPython tutorial.
\end{itemize} 

\item {\bf Run using  MPICH2 (1.0.5):}  
\begin{itemize}
\item {\tt Baseline} is a ``hello world'' type application included to show the cost of checkpointing MPICH2 and its resource manger, MPD. 
\item {\tt ParGeant4:} Geant4~\cite{Geant4} is a million-line C++ toolkit for
  simulating particle-mattter interaction.  It is based at CERN, where
  the largest particle collider in the world has been built.
  ParGeant4~\cite{AlversonEtAl01} is a parallelization based on TOP-C, that is distributed
  with the Geant4 distribution.  TOP-C (Task Oriented Parallel C/C++)
  was in turn built on top of MPICH2 for this demonstration.
\item {\tt NAS NPB2.4:} CG (Conjugate Gradient, level~C) from the well-known 
  benchmark suite NPB.  NPB~2.4-MPI was used.
\end{itemize}

\item {\bf Run using OpenMPI (1.2.4):}
\begin{itemize}
\item {\tt Baseline} is a ``hello world'' type application included to show the cost of checkpointing OpenMPI and its resource manger, OpenRTE. 
\item {\tt NAS NPB2.4:} a series of well-known MPI benchmarks.
  NPB~2.4-MPI was used.  The benchmarks run under OpenMPI are: 
  BT (Block Tridiagonal, level~C: 36~processes since the software requires a square number),
  SP (Scalar Pentadiagonal, level~C: 36~processes since the software requires a square number),
  EP (Embarrassingly Parallel, level~C),
  LU (Lower-Upper Symmetric Gauss-Seidel, level~C),
  MG (Multi~Grid, level~C), and
  IS (Integer Sort, Level~C).
\end{itemize}

\end{itemize}

In Figure~\ref{fig:nodespg4local} we use ParGeant4 as a test case to
report on scalability with respect to the number of nodes.  When resource  
management processes are included, we are checkpointing a total of 289
processes in the largest example. Figure~\ref{fig:nodespg4san} repeats
this tests with checkpoints written to a centralized storage device.

\begin{figure}[htbp]
  \begin{center}
  \resizebox{.5\linewidth}{!}
  {\includegraphics{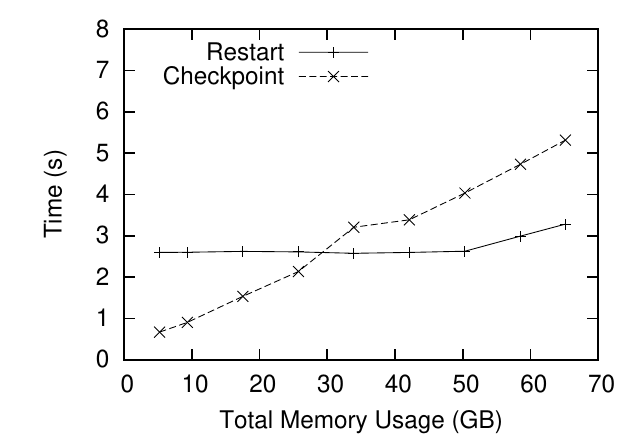}}
  \end{center}
  \caption{\label{fig:lgm_local}
  Timings as memory usage grows.  A synthetic OpenMPI program allocating
  random data on 32 nodes.  Compression is disabled.  Checkpoints
  written to local disk.  }
\end{figure}

Figure~\ref{fig:lgm_local} illustrates the time as memory
usage grows, while holding fixed the number of participating nodes
at~32.  The implied bandwidth is well beyond the typical 100~MB/s
of disk, and is presumably indicating the use of secondary storage
cache in the Linux kernel.  Restart times also indicate the use of cache
and page table optimizations in the kernel.

An optional feature in DMTCP is to issue a {\tt sync} after checkpointing
to wait for kernel write buffers to empty before resuming the user threads.
Results shown do not issue a call to {\tt sync}. This is consistent with
timing methodology most prevalent in related work.  The cost of issuing a
{\tt sync} can be easily estimated based on checkpoint size and disk speed.
As an example, if a {\tt sync} is issued for ParGeant4 (compression enabled)
a mean additional cost of 0.79 seconds (with a standard deviation of 0.24)
is incurred. An alternate strategy is to sync the {\em previous} checkpoint
instead.  This has the benifits of still guaranteeing the consistency of
all except the last checkpoint without having to wait for disk in most cases.

\subsection{Breakdown of Checkpoint/Restart Times}

\begin{table}[ht]
\centering
\subfloat[Checkpoint\label{tab:ckptstates}]{
\begin{tabular}{|l|r|r|r|}
\hline
 Stage                  & Uncompressed & Compressed & Fork Compr. \\
\hline
Suspend user threads    & 0.0251   & 0.0217     & 0.0250  \\
Elect FD leaders        & 0.0014   & 0.0013     & 0.0013  \\
Drain kernel buffers    & 0.1019   & 0.1020     & 0.1017  \\
Write checkpoint        & 0.6333   & 3.9403     & 0.0618  \\
Refill kernel buffers   & 0.0006   & 0.0008     & 0.0016  \\
\hline                                          
Total                   & 0.7630   & 4.0669     & 0.1922  \\
\hline
\end{tabular}}

\subfloat[Restart\label{tab:rstrstates}]{
\begin{tabular}{|l|r|r|}
\hline
Stage & Uncompressed & Compressed  \\
\hline
Restore files and ptys    & 0.0056 & 0.0088 \\
Reconnect sockets         & 0.0400 & 0.0214 \\
Restore memory/threads    & 0.8139 & 2.1167 \\
Refill kernel buffers     & 0.0009 & 0.0018 \\
\hline
Total                     & 0.8604 & 2.1487 \\
\hline
\end{tabular}}
\caption{\label{tab:stages}Time (in seconds) for different stages of checkpoint
and restart for NAS/MG under OpenMPI, using 8~nodes.  Forked is the same as
compressed, except that compression and writing are delegated to a child
process and allowed to run in parallel.  
}
\end{table}

Table~\ref{tab:stages} shows the times for the different stages of
checkpointing and restart.  Checkpoint time is dominated by writing the
checkpoint image to disk.  This timing breakdown is typical for all
other applications we examined.  The time for writing the checkpoint
image to disk is almost entirely eliminated by using the technique of
forked checkpointing~\cite{ForkedCheckpointing90,ForkedCheckpointing94}.
In forked checkpointing, a child process is forked, the child process writes
the checkpoint image, and the parent process continues to execute, taking
advantage of UNIX copy-on-write-semantics.  Forked checkpointing has the
disadvantage that compression runs in parallel and may slow down the user
process and take longer.  The forked checkpointing times presented here
are from an experimental version of DMTCP.  Forked checkpointing was also
supported in a forerunner of the current work~\cite{CoopermanAnselMa06}.

The stages in Table~\ref{tab:ckptstates} correspond to steps 2~through 6~in
Figure~\ref{fig:ckptdiag}. The stages in Table~\ref{tab:rstrstates} correspond
to steps 1~through 6~in Figure~\ref{fig:rstrdiag}, except that steps 3 and
4, which take negligible time, are lumped in with step 5.  Since the first
3 reported times for restart occur in parallel on each node, the reported
times are an average across all 8 nodes.  All other times are the durations
between the global barriers.

\subsection{Experimental Analysis}

In principle, the time for checkpointing is dominated by:
(i)~compression (when enabled); (ii)~checkpointing memory to disk; and
(iii)~(to a much lesser extent) flushing network data in transit and re-sending.  When
compression is enabled, that time dominates.  The cost of
flushing and re-sending is bounded above by the size of the
corresponding kernel buffers and the capacity of the network switches,
which tend to be on the order of tens of kilobytes.
Restart tends to be faster than checkpoint, because {\tt gunzip} operates
more quickly than {\tt gzip}.

The graphs in Figure~\ref{fig:apps32} show that the time to checkpoint
using compression tends to be slowest when the uncompressed checkpoint
image is largest.  An exception occurs for NAS/IS.  NAS/IS is a
parallel integer bucket-sort.  The bucket sort code has allocated
large buckets to guard against overflow.  Presumably, the unwritten
portion of the bucket is likely to be mostly zeroes, and it compresses
both quickly and efficiently.

Figure~\ref{fig:nodespg4} shows the time for checkpoint and
restart to be insensitive to the number of nodes being used.  This is
to be expected since checkpointing on each node occurs asynchronously.
It also demonstrates that the single checkpoint coordinator, which
implements barriers, is not a bottleneck.  In the event that it were a
bottleneck, we would replace it by a distributed coordinator in our
implementation.  

\section{Conclusions and Future Work}
\label{sec:conclusion}

A scalable approach to transparent distributed checkpointing has been
demonstrated that does not depend on a specific message passing library.
Nor does it depend on kernel modification.  The approach achieves broad
application coverage across a wide array both of scientific and common
desktop applications.  On 128~distributed cores (32~nodes), a typical
checkpoint time is 2~seconds, or 0.2~seconds by using forked checkpointing,
along with negligible run-time overhead.  This makes DMTCP attractive both
for frequent checkpointing and for minimal application interruptions during
checkpointing of interactive applications.  Experimental results have shown
that the approach is scalable and that timings remain nearly constant as nodes
are added to a computation within a medium-size cluster.  The centralized
checkpoint coordinator, which implements barriers, has minimal overhead
in these experiments.  As the approach is scaled to ever larger clusters,
the single coordinator can be replaced by a distributed coordinator using
well-known algorithms for distributed global barriers and distributed
discovery services.

In the future, it is hoped to support new communication models such as
multicast and RDMA (remote direct memory access) as used in networks such
as InfiniBand.  Future work will fully support the ptrace system call,
and therefore checkpointing of gdb sessions.  Future work will also
extend the ability to checkpoint X-Windows applications, as currently
demonstrated by the simple checkpointing of TightVNC.  This will further
enhance the attractiveness of user-level checkpointing.

\section{Acknowledgements}

We thank our colleagues at CERN who have discussed, helped debug, and tested
the use of DMTCP on runCMS and on ParGeant4.  In particular, we thank John
Apostolakis, Giulio Eulisse and Lassi Tuura.  We thank Xin Dong for his help
in installing and testing ParGeant4 in a variety of operating circumstances.
We acknowledge the gracious help of Michael Rieker in numerous discussions,
and in fixing more than one bug in MTCP for us.  We also acknowledge the help
of Daniel Kunkle in testing checkpointing of TightVNC.  Finally, we thank
Peter Keller, David Wentzlaff, and Marek Olszewski for helpful comments on
draft manuscripts.

\bibliographystyle{plain}
\bibliography{dmtcp}

\end{document}